\documentclass[final,3p,times,twocolumn]{elsarticle}
\usepackage{amsmath}
\biboptions{sort&compress}
\usepackage{float}
\usepackage{comment}

\usepackage[unicode=true,
 bookmarks=true,bookmarksnumbered=true,bookmarksopen=true,bookmarksopenlevel=1,
 breaklinks=true,pdfborder={0 0 1},backref=false,colorlinks=true]
 {hyperref}
\hypersetup{pdftitle={Gapped-filtering},
 pdfauthor={Minh Nguyen},
 linkcolor=black,citecolor=black,urlcolor=blue,filecolor=blue,pdfpagelayout=OneColumn,pdfnewwindow=true,pdfstartview=XYZ,plainpages=false}

\makeatletter
\@ifundefined{date}{}{\date{\today}}
\usepackage{babel}
\usepackage{chemformula}
\usepackage{braket}

\makeatother

\begin{document}
\title{Gapped-filtering for efficient Chebyshev expansion of the density projection operator}

\author{Minh Nguyen and Daniel Neuhauser}
\affiliation{Department of Chemistry and Biochemistry, University of California at Los Angeles, Los Angeles, California 90095, USA}
\begin{abstract}
We develop the gapped-filtering method, whereby a short Chebyshev expansion accurately represents the density-matrix operator. 
The method optimizes the Chebyshev coefficients to give the correct density matrix at all energies except within the gapped region where there are no eigenstates. 
Gapped filtering reduces the number of required terms in the Chebyshev expansion compared to traditional expansion methods, as long as one knows or can determine efficiently the HOMO and LUMO positions. 
The reduction is especially noticeable (factors of 2-3) when high accuracy is sought.
To exemplify the method, we use gapped-filtering to increase the efficiency of stochastic-GW calculations.
\end{abstract}
\maketitle
\section{Introduction}
A fundamental ingredient in many electronic structure and dynamics methodologies is the polynomial expansion of the density matrix operator.
Such methods include deterministic approaches to DFT that avoid the direct diagonalization of the Hamiltonian matrix,~\citep{baer1997chebyshev,jay1999electronic} or allow for fast diagonalization of the Kohn-Sham Hamiltonian~\cite{liang2003improved} when the Hamiltonian is a sparse matrix. 
These polynomial expansions are also the primary ingredient in stochastic quantum chemistry methods,~\citep{sankey1994projected} including  stochastic DFT~\cite{baer2013self,neuhauser2014fragment,baer2022stochastic} and beyond-DFT approaches. 
The latter include the linear-scaling stochastic-GW method (sGW)~\cite{neuhauser2014breaking} which calculates the quasiparticle (QP) energy as a perturbative correction to the DFT eigenvalue~\cite{friedrich2006many,hybertsen1986electron,vlcek2021stochastic,vlvcek2018swift}, as well as, e.g.,  stochastic-MP2~\cite{neuhauser2013expeditious} stochastic Bethe Salpeter Equation~\citep{bradbury2022bethe} and stochastic Time-Dependent DFT (TDDFT)~\citep{gao2015sublinear}.

In the usual Chebyshev approach \cite{TalEzer1984,kosloff1988time,chebyshev2006RMP} the coefficients in the polynomial expansion of the density matrix $\hat{\rho}$ are obtained analytically by requiring that the scalar function $\Theta(\mu-E)$ converges uniformly as a function of $E$~\citep{kosloff1988time,chebyshev2006RMP}.
In practice, to converge the expansion one usually replaces the sharp Heaviside function with a smoothed one.
However, for most systems the gap is small relative to the full energy range of the Hamiltonian operator, so even a smoothed Heaviside function would require typically thousands of terms for convergence.

Here we suggest an alternative to the usual determination of the Chebyshev coefficients.
For any desired number of coefficients, we only require that the Chebyshev expansion be correct for energies outside the gap.  
The logic is that it is immaterial what the values of the polynomial expansion are inside the gap since there are no eigenstates there. 
Within the gap the filter could have any form, including Gibbs oscillations, but they would be irrelevant to the final density matrix. 

In Section II we develop the idea in detail and give a simple prescription, which we label gapped-filtering, for obtaining the Chebyshev coefficients.  
Numerically, gapped-filtering carries negligible overhead.  
One just specifies the desired length of the Chebyshev expansion, labeled $N_{\rm chb}$, and then just inverts a single matrix of rank $N_{\rm chb}+1$.  

In Section III we numerically study gapped-filtering and show that the method converges rapidly with $N_{\rm chb}$, which is the only parameter in the approach. 
The method is more efficient than traditional smoothed filters, where one needs to converge with both a smoothness parameter and $N_{\rm chb}$. When high accuracy is required, gapped filtering is 2-3 times faster than the traditional approaches.

Section IV presents an application of the method to stochastic-GW, verifying this significant reduction in a large scale calculation. Conclusions are discussed in Section V.

\section{Methodology}
\subsection{Gapped-filtering}
Ignoring spin, the density matrix, i.e., the projection operator to the occupied manifold, is formally:
\begin{equation}
    \hat{\rho} = \sum_{i=1}^{N_{\rm{occ}}} \ket{\phi_i}\bra{\phi_i},
\end{equation}
where $\{\phi_i\}$ are the eigenvectors of the 1-body Hamiltonian operator $\hat{H}$ and $N_{\rm{occ}}$ is the number of occupied states. 
We only consider here systems with a gap where the physical temperature is tiny compared to the gap, i.e., a zero-temperature description.
 
Even if all eigenstates and eigenvalues are known, the memory required to store the eigenstates could be enormous for giant systems with tens of thousands of electrons. In those cases, it is better to use a polynomial expansion. 
Thus, the density matrix is equivalent to a Heaviside function of energy centered within the bandgap,
\begin{equation}
    \label{eq:density matrix theta}
    \hat{\rho} = \Theta(\mu-\hat{H}),
\end{equation}
where $\mu$ is the chemical potential. 
To converge the expansion, a smoothed complementary-error function is typically used:
\begin{equation}   
  \Theta(\mu-\hat{H}) \simeq \hat{\Theta}_\beta \equiv \sqrt{\frac{1}{2}{\rm{erfc}}\left(\beta\left(\mu-\hat{H}\right)\right)},
  \label{eq:ThetaErfc}
\end{equation}
where $\beta$ is an inverse-temperature-like parameter that determines the sharpness of the function~\cite{liang2003improved,baer1997chebyshev,jay1999electronic}.  

Note that the square root stems from the fact that in most cases, especially for stochastic quantum chemistry applications, the results of applying the filter are then squared to give the density or Green's function. 
Throughout the paper, the analytical filters always refer to this square-root erfc form (abbreviated  as sqrt-erfc).  
Further, we verified that gapped-filtering  does analogously well when compared with an erfc filter without a square-root.

To apply the filter, the Chebyshev polynomial expansion is traditionally used:
\begin{equation}  
    \Theta(\mu-\hat{H}) \simeq \sum_{n=0}^{{N_{\rm{chb}}}}a_n T_n(\Tilde{H}),
    \label{eq:ChebExp}
\end{equation}
where the expansion coefficients, $a_n$,  depend on $\mu$ and $\beta$ and $T_n$ which is the n'th order Chebyshev polynomial of the first kind.
The argument of the Chebyshev function is a scaled Hamiltonian, $\Tilde{H}\equiv \frac{{\hat{H}}-H_{\rm{avg}}}{\Delta H}$,~\cite{kosloff1988time,mandelshtam1993calculation} where the $H_{\rm{avg}}$ and $\Delta H$ parameters are the center and half-width of the spectrum of $\hat{H}$, so that the eigenvalues of $\Tilde{H}$ are  between -1 and 1.
The length of the expansion, ${N_{\rm{chb}}}$, is approximately proportional to
$\beta \Delta H$, times a factor which depends on the relative position of $\mu$ relative to the spectrum bottom, $H_{\rm avg}-\Delta H$. (Qualitatively that factor accounts for the ``stretching" of the angle $\theta=\arccos(x)$ near the band bottom, where $x\to -1$ and $dx/d\theta$ becomes large.  For a fuller discussion see \cite{Suryanarayana2013}.)

Usually the coefficients are determined by requiring that the expansion in Eq. (\ref{eq:ChebExp}) is valid uniformly over all $E$ between the minimum and maximum eigenvalues of $\hat{H}$. 
Due to the orthogonality of the Chebyshev polynomials, the coefficients are then simply:
\begin{equation}
    a_n=\frac{2-\delta_{n0}}{\pi} \int_{-1}^{1} \Theta(\tilde{\mu}-x)
    T_n(x) w(x) dx, \label{eq:classicalcoefs}
\end{equation}
where $w(x)=(1-x^{2})^{-1/2}$ is the Chebyshev weighting function and we introduced the scaled chemical potential, $\tilde{\mu}=(\mu-H_{\rm avg})/\Delta H$.

Here we exploit a simple realization: because there are no states within the bandgap, the behavior of the expansion within the gap does not impact the density matrix.
What is important is that the expansion is 1 for the occupied states and 0 for the unoccupied states, regardless of its values within the bandgap. 
As we show below, the increased flexibility, as the expansion is free to vary within the bandgap, suffices to reduce the number of polynomials for a given desired level of numerical accuracy.

Our derivation starts by defining a modified weighting function that vanishes within the scaled bandgap:
\begin{equation}
    w^{*}(x)=w(x)\big(\Theta(x_H-x)+\Theta(x-x_L)\big),
\end{equation}
where the scaled energies are
$$x_H=(\varepsilon_{{\rm {H}}}-H_{\rm{avg}})/\Delta H+\delta$$  $$x_L=(\varepsilon_{{\rm {L}}}-H_{\rm{avg}})/\Delta H-\delta$$
where $\varepsilon_H$ and $\varepsilon_L$ are the HOMO and LUMO energies, while $\delta$ is a small padding factor, which we usually take to be about $1\%$ of the scaled bandgap.
This padding ensures that the weighting function does not go to 1/0 at exactly the HOMO/LUMO energies, thereby allowing for any uncertainties in the values of the HOMO/LUMO energies.

Since the weighting function is zero within the gap, we call the resulting method gapped-filtering, and the associated Chebyshev expansion gapped-filter.  As the Chebyshev polynomials are not orthogonal under this weighting function, the coefficients need to be rederived.
This is done by minimizing the norm squared of the difference between the Heaviside function and the Chebyshev expansion, weighted now by $w^{*}(x)$:
\begin{equation}
    J=\int_{-1}^{1}\bigg|\Theta(x-x_{0})-\sum_{n=0}^{{N_{\rm{chb}}}}a_{n}T_{n}(x)\bigg|^{2}w^{*}(x)dx.
\end{equation}
Minimizing $J$ with respect to $a_{i}^*$ gives:
\begin{equation}
    \frac{\partial J}{\partial a_{i}^{*}}=\sum_{k}M_{ij}a_{j}-b_{j}=0,
\end{equation}
i.e., $a=M^{-1} b$, 
where $$b_{i}=\int_{-1}^{x_H} w^*(x) T_i(x) dx,$$
and 
\begin{equation}M_{ij}=\int_{-1}^1 w^*(x)T_i(x)T_j(x) dx.
 \label{eq:coefficient_matrix}
 \end{equation}

\begin{figure}[H]
    \centering
    \includegraphics[width=0.95\columnwidth]{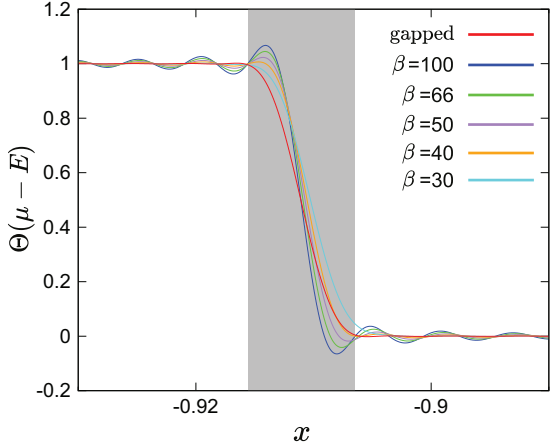}
    \caption{Gapped-filtering vs.~various sqrt-erfc filters, using a Chebyshev expansion with  ${N_{\rm{chb}}=450}$ polynomials, as a function of the scaled energy, $x=(E-H_{\rm avg})/\Delta H$.  Naphthalene parameters are used ($\Delta H= 13.5$~Hartree, $\varepsilon_{\rm gap}=0.123$~Hartree). The highlighted grey area denotes the gap region where the expansion need not be optimized. Note that $\beta$ is always reported in inverse Hartree, so, for example, for  $\beta=30$~Hartree$^{-1}$ the product of $\beta$ with the gap is quite small, $\beta (\varepsilon_L-\varepsilon_H)=3.7$. }
    \label{fig:1_expansion}
\end{figure}

The matrix elements are easily calculated.  Defining $x=\cos\theta$, we get:
\begin{equation}
    \label{eq:M matrix}
    M_{ij}=F_{ij}(-1,x_H)+F_{ij}(x_L,1),
\end{equation}
where 
\begin{equation}
    \begin{aligned}
     F_{ij}(y,z) & =\int_{y}^{z}w(x)T_{i}(x)T_{j}(x)\\
        & =\int_{\theta_z}^{\theta_y}\cos(i\theta)\cos(j\theta)d\theta\\
        & =\frac{1}{2}\int_{\theta_{z}}^{\theta_{y}}\big(\cos((i-j)\theta)+\cos((i+j)\theta)\big)d\theta\\
        & =G_{ij}(\theta_{y})-G_{ij}(\theta_{z})
    \end{aligned},
\end{equation}
where $\cos\theta_{y}=y$, and 
\begin{equation}
    G_{ij}(\theta)=
    \begin{cases}
        \theta & i=j=0\\
        \frac{1}{2}\theta+\frac{1}{4i}\sin(2i\theta) & i=j\neq0\\
        \frac{\sin((i-j)\theta)}{2(i-j)}+\frac{\sin((i+j)\theta)}{2(i+j)} & i\neq j
    \end{cases}.
\end{equation}
Similarly, 
\begin{equation}
    b_{i}=\int_{\theta_H}^{\pi}\cos(i\theta) d\theta=
    \begin{cases}
        \pi-\theta_{H} & i=0\\
        -\frac{1}{i}\sin(i\theta_{H}) & i>0
    \end{cases},
\end{equation}
where $\cos(\theta_{H})=x_H$.

\subsubsection*{Additional considerations}

We now discuss several issues associated with the method.

\begin{figure}[H]
    \centering
    \includegraphics[width=0.95\columnwidth]{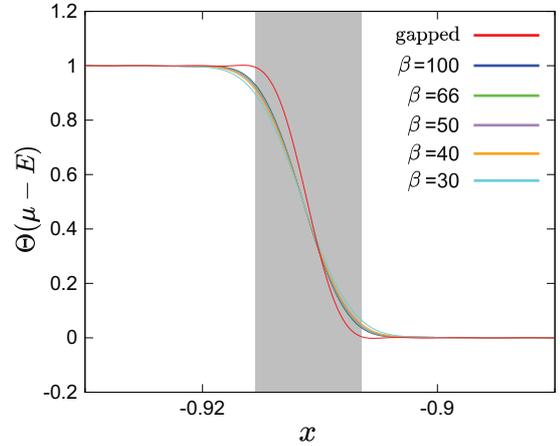}
    \caption{Analogous to Fig. 1 but now the analytical coefficients are damped using the commonly used Jackson kernel coefficients to avoid a sharp cut at $N_{\rm chb}$.
    While the Gibbs oscillations are damped the step function is even further widened, i.e., gets further from the analytical step function demonstrating that just damping the analytical coefficients is not sufficient.}
    \label{fig:2_expansion_KPM}
\end{figure}

\textbf{HOMO/LUMO determination:} Our method requires knowledge of the HOMO/LUMO energies associated with the one-body Hamiltonian $\hat{H}$.  
In situations where these are not accurately given, it will often be worthwhile to do a single longer Chebyshev propagation to determine accurately, by filtering, the location of the HOMO and LUMO.  
For example, in stochastic GW the gapped filter will be applied thousands of times, so the initial overhead for finding the HOMO/LUMO energies by a narrow Chebyshev filter will be often small compared to the savings incurred by the subsequent use of gapped filtering in sGW. 

The same considerations would also be true in DFT applications that rely on Chebyshev filtering, as long as the overall number of vectors on which one needs to apply Chebyshev filtering is much larger than 1, so that in each SCF step the extra overhead in a single long Chebyshev to determine the HOMO and LUMO would be negligible compared to the overall cost needed to filter all the required states.

\textbf{HOMO/LUMO accuracy:} Note that formally the HOMO and LUMO energies that are used need to be lower and upper bounds to the true HOMO and LUMO, but since the method is continuous even a small (circa 5$\%$ or less) deviation in the wrong direction (e.g., an applied HOMO energy which is higher than the true HOMO by about 2-3$\%$ of the gap, and analogously for the LUMO) still gives excellent results, as we verified.

\textbf{Correlation function:} Also note that when one is just looking for a correlation function, involving a matrix element of the density of state (rather than the full filtered function, Eq. (\ref{eq:ThetaErfc})), then the total number of residues is halved by using simple trigonometric identities, e.g., $2\braket{\zeta_0 |T_{2n}   |  \zeta_0}     = 
\braket{\zeta_n |\zeta_n } - \braket{ \zeta_0 |\zeta_0 }$ where $\ket{\zeta_n} \equiv T_n \ket{\zeta_0} $ (see Eq. 3.10 in \cite{Wall1995}). 
This reduction by a half is valid for both the regular Chebyshev approaches and for gapped filtering, so it does not change the relative performance of the approaches.    

\textbf{Filter design:} To conclude this section, note that the problem we specifically try to solve, an optimum Chebyshev filter where we do not care  what happens inside the gap, could be viewed as a specific sub-problem of digital filter design   \cite{DigitalFilters1973,parks_burrus_1987}.

\section{Gapped-filtering convergence study}
We now turn to a numerical study of the new method. 
We use the band gap and energy width from an LDA calculation of Naphthalene.  
The calculations are detailed in the next section, but the relevant part here is that the gap is $\varepsilon_{\rm gap}\equiv \varepsilon_L-\varepsilon_H=0.123$~Hartree, the spectrum width is $13.5$~Hartree, and the scaled HOMO/LUMO energies are $(\varepsilon_H-H_{\rm avg})/\Delta H =-0.9156$ and $(\varepsilon_L-H_{\rm avg})/\Delta H=-0.9065$ 
i.e., a scaled gap of $(\varepsilon_L-\varepsilon_H)/\Delta H = 0.0091$.

\begin{figure}[H]
    \centering
    \includegraphics[width=0.95\columnwidth]{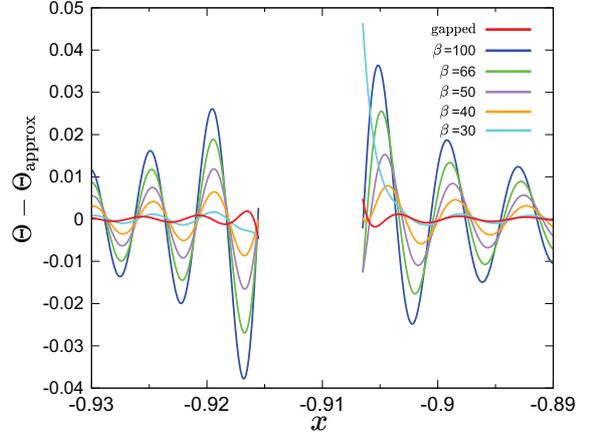}
    \caption{Analogous to Fig. \ref{fig:1_expansion}, but showing, away from the gap, the differences between the Chebyshev expansions and an exact step function.}
    \label{fig:3_diff_expansions}
\end{figure}

In Fig. \ref{fig:1_expansion} we show, for an expansion length of $N_{\rm chb}=450$, the improved performance of gapped filter over the traditional approaches.  We specifically plot as a function of the scaled energy ($x=(E-H_{\rm avg})/\Delta H)$, the numerical representation of  $\Theta(\mu-E)$ (i.e., the RHS of Eq. (\ref{eq:ChebExp})) at several $\beta$ values.  As expected, convergence is faster for a lower $\beta$, but if $\beta$ is too low then the analytic sqrt-erfc function deviates too much from the desired step function.

As a side note, a common approach   \cite{Weie2006}  to avoid Gibbs oscillations is to multiply the Chebyshev coefficients by a filter, $c_n\to g_n c_n $, where $g_n$ falls smoothly to almost 0 at $n=N$.  This is shown in Fig. \ref{fig:2_expansion_KPM} (constructed for $N_{\rm chb}=450$, like Fig. \ref{fig:1_expansion}), using the popular Jackson filter \citep{Weie2006}. But while the smoothing of the analytic coefficients does remove the Gibbs oscillations, this comes however at the expense of making the filter wider, so it is even further from the true step function filter. 

The difference in Fig. \ref{fig:1_expansion} at  $N_{\rm chb}=450$ between the numerical representations and the desired step function is detailed in Fig. \ref{fig:3_diff_expansions}. Interestingly, at this $N_{\rm chb}$ the gapped filter is also smooth inside the gap (gray region in Fig. \ref{fig:1_expansion}) even though the gapped-filtering method does not try to explicitly optimize the behaviour of the filter there.  
(Note: for very high $N_{\rm chb}$ which are irrelevant, $N_{\rm chb}>1300$, gapped-filtering does show Gibbs oscillations inside the gap.)

\begin{figure}[H]
    \centering
    \includegraphics[width=0.95\columnwidth]{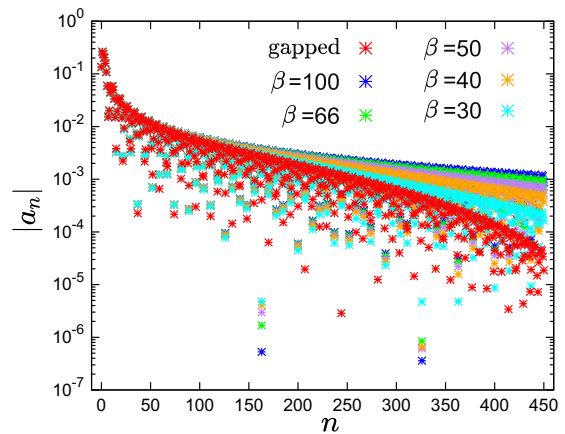}
    \caption{The absolute value of the expansion coefficients for $N_{\rm{chb}} = 450$ for the gapped and traditional (sqrt-erfc) filters.}
    \label{fig:4_abs(co)}
\end{figure}

In Fig. \ref{fig:4_abs(co)} we show, again at $N_{\rm chb}=450$, the expansion coefficients, $a_n$, as a function of $n$. The magnitude of the gapped-filtering coefficients decreases at a much faster rate than that of the traditional coefficients.  The faster drop-off leads to a smoother expansion and therefore to less oscillations.

We now turn to an analysis of the performance of the gapped-filtering approach as a function of the number of Chebyshev polynomials.  Fig.~\ref{fig:5_max_diff} shows, as a function of  $N_{\rm{chb}}$, the maximum (over all energies outside the gap) of the absolute difference between Eq.~(\ref{eq:ChebExp}) and Eq.~(\ref{eq:density matrix theta}) at several $\beta$ values. The relative advantage of gapped filtering is more pronounced when higher accuracy is sought, i.e., at larger $N_{\rm chb}$.  Thus small $\beta$ values (in our case $\beta=30$ or $40$~Hartree$^{-1}$) require a small number of Chebyshev polynomials for convergence, that rivals the number in gapped-filtering. However, for such small $\beta$ values the analytical sqrt-error function deviates significantly from the desired step function, so that the error would be high at any value of $N_{\rm{chb}}$. In contrast, a more accurate representation of the step function with small error requires a much larger number of polynomials for sqrt-erfc filters (see here the graphs for  $\beta=50,66,100$~Hartree$^{-1}$) than in gapped filtering.    

A technical point is that for very large $N_{\rm chb}$, some of eigenvalues of $M$ will be close to zero; this would lead to round-off errors in the matrix inversion, which in our case would deteriorate the accuracy of the gapped filter when the maximum error is on the order of $10^{-6}$, i.e., beyond $N_{\rm{chb}}> 10 \Delta H/ \varepsilon_{\rm gap} = 1100$.  
We avoid such errors using, for such large $N_{\rm chb}$, quadruple precision in calculating the coefficients (i.e., for yielding $M$ and $b$, and then for inverting $M^{-1} b$ using a quadruple precision  algorithm~\citep{dongarra1979linpack,lawson1979basic}).
Because the dimensions of $M_{ij}$ are 
$(N_{\rm{chb}}+1)\times ( N_{\rm {chb}}+1)$ and 
${ N_{{\rm {chb}}}}$ is typically on the order of $100-2000$, this inversion has negligible cost. (To clarify, the coefficients are then well-behaved and are used in the usual double-precision Chebyshev algorithm, i.e., quadruple precision is only used to determine the $a_n$ coefficients.)

An advantage of gapped-filtering is that it avoids the use of an artificial temperature parameter, $\beta$.  We only need to worry about a single convergence parameter, $N_{\rm{chb}}$, unlike the usual procedure of converging with respect to two parameters, $\beta$ and $N_{\rm{chb}}$.

\section{Application: stochastic-GW}
\subsection{Filtering in stochastic-GW}
The polynomial expansion of the Chebyshev operator is crucial for  our stochastic quantum chemistry approaches.  
Here we overview the relevant parts for one such method, stochastic-GW.  
Only the parts of the method where filtering is relevant are discussed, see Refs.~\citep{neuhauser2014breaking,vlvcek2018swift} for further details.
The stochastic-GW method, in its simplest version, calculates the HOMO (or LUMO) quasiparticle energy in the diagonal approximation $\varepsilon_{QP}$, i.e., based on the HOMO/LUMO DFT eigenvector whose associated eigenvalue is modified.

Stochastic-GW has two filtering stages.  
First, one does a stochastic realization of $G$. 
For each such realization, one takes an initially random white noise orbital $\ket{\zeta_0},$ i.e., $\zeta_0(r)\propto \pm 1$ in a grid representation (or an analogous expression in a basis-set description).  Then, the white noise orbital is filtered with the Heaviside operator $\hat{\Theta}\equiv \Theta(\mu-\hat{H})$ to generate  a function $\ket{\zeta}$ which contains only occupied states (with random coefficients):
\begin{equation}
    \label{eq:filtered_orb xi}
    \zeta(r)=\bra{r} \hat{\Theta} \ket{\zeta_0}.
\end{equation}
We interchangeably label such a function as a stochastic occupied orbital, or more simply a filtered orbital.

Each such filtered $\zeta$ function is then propagated in time under $\hat{H}$; the correlation function of the result with $\zeta_0$ equals the negative-time Green's function, $\{\zeta(r,t) \zeta_0(r')\}= G_0(r,r',-t)$.  The positive-time Green's function is obtained analogously.  The curly brackets indicate a statistical average over the stochastic orbitals. Typically we use $N_\zeta=$ 200 to 2000  different random white noise $\zeta_0(r)$ functions.  Fewer samples are needed for larger systems due to self-averaging.

Then, for each such stochastic realization of $G$ (i.e., for each $\zeta_0$), one calculates the action of the time-dependent effective interaction ${\hat{W}}(t)$ on a vector related to $\zeta_0$.  This is done by choosing, for each $\zeta_0$, several ($N_{\rm{orb}}$) white noise vectors, labeled $\eta_{0\ell}(r),\,\,\,\ell=1,...,N_{\rm orb}$.  These vectors are also filtered, $|\eta_\ell\rangle=\Theta |\eta_{0\ell}\rangle$.  
This set of filtered orbitals are then excited and propagated in time, under a time-dependent Hartree Hamiltonian $H(t)$. 
The crucial part of the evolution Hamiltonian, which also gives the action of $\hat{W}(t)$, comes from the Hartree potential due to the time-dependent density  
\begin{equation}
    n(r,t)\simeq\frac{2}{N_{\rm{orb}}}\sum_{\ell=1}^{N_{\rm{orb}}}|\eta_\ell(r,t)|^2,  
\end{equation}
where $\eta_\ell(r,t=0)$ is obtained by slightly "kicking" the filtered orbital $\eta_\ell(r)$ (for further details 
see Refs.~\cite{neuhauser2013expeditious,gao2015sublinear}).  Note that we consider closed shell systems so the factor of 2 due to spin was restored here. 

The sGW formalism is efficient since very few orbitals (typically $N_{\rm orb}=10$, and less for larger systems) are needed to describe the response of molecules and solid-state systems within the short few-fs time for which $\hat{W}(t)$ is needed.  Still, a total of $N_\zeta N_{\rm orb} \simeq 2000-40,000$ filtering projections is required, therefore motivating the use of the proposed gapped-filtering approach.
    
\subsection{Gapped-filtering for stochastic-GW}
We now demonstrate gapped-filtering in an sGW study of naphthalene.
All calculations use $N_\zeta=1000$ stochastic samplings of $G$, which gives the quasi-particle energy to within a statistical error of $\pm0.06$~eV.
 
The DFT Hamiltonian $\hat{H}$ uses the LDA exchange-correlation functional with Troullier-Martins norm-conserving pseudo-potentials~\cite{hamann2013optimized}. The grid employed has $48 \times  44 \times 24$ points with a grid spacing of 0.5~Bohr. As mentioned, the DFT bandgap for naphthalene  is then $\varepsilon_{\rm gap} = 3.34$~eV, i.e., $0.123$~Hartree, while the spectrum half-width $\Delta H$ is 13.5~Hartree, i.e., 110 times higher than the bandgap.
\begin{figure}[H]
    \centering
    \includegraphics[width=0.95\columnwidth]{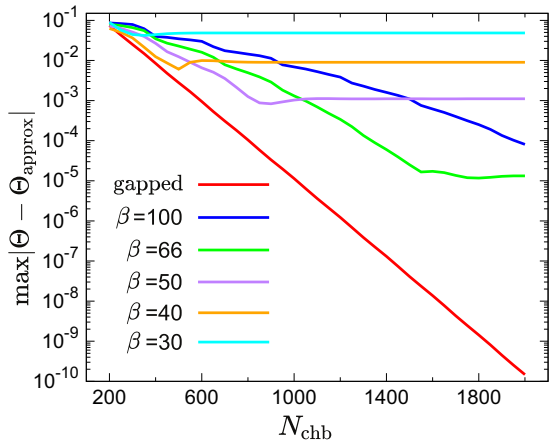}
    \caption{The maximum, taken over all energies outside the bandgap, of the absolute value of the difference between the step function $\Theta(\mu -E)$ and its Chebyshev expansion (the RHS of Eq. (\ref{eq:ChebExp})).  
    The figure is plotted as a function of the number of Chebyshev terms, $N_{\rm{chb}}$.  Note that smoother low-$\beta$ filters converge initially faster but then their error reaches quickly a high plateau.}
    \label{fig:5_max_diff}
\end{figure}

\begin{figure}[H]
    \centering
    \includegraphics[width=0.95\columnwidth]{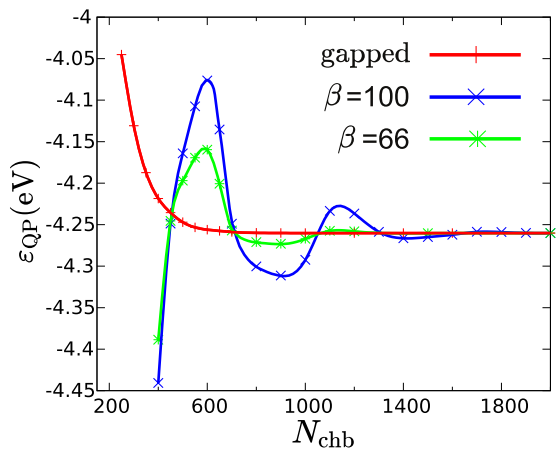}
    \caption{The stochastic-GW quasiparticle energy of naphthalene, $\varepsilon_{\rm {QP}}$, as a function of the number of Chebyshev polynomials, ${N_{\rm{chb}}}$, using  gapped-filtering and sqrt-erfc filters with $\beta=66$ and $\beta=100$.
    Gapped-filtering converges to the asymptotic QP energy at much lower $N_{\rm{chb}}$ than the traditional sqrt-erfc filters.}
    \label{fig:6_QP_v_nchb}
\end{figure}

Fig.~\ref{fig:6_QP_v_nchb} shows the resulting HOMO QP energies  for gapped-filtering  and for the traditional filtering method from Eq. (\ref{eq:ChebExp}) at $\beta=66$ and $\beta=100$.
The statistical errors barely change with $N_{\rm chb}$, so we do not show them in the figure.

The figure shows that gapped-filtering requires very few Chebyshev terms.  Thus, to have a systematic error of 0.015~eV or better, $N_{\rm chb}=450$ is sufficient for gapped-filtering.

\section{Conclusions}
In conclusion, our work has shown that, for gapped systems,  a step-function filter in energy is best done by optimizing the Chebyshev coefficients to fit the gapped spectrum only, rather than using a smooth filter and unnecessarily fitting the irrelevant gapped region.

The final algorithm is very easy to adopt for practical systems.  Given the HOMO/LUMO energies and the spectrum overall min/max energies, one should specify an acceptable tolerance of the deviation from the step function, and then do a similar calculation the red line in Fig. \ref{fig:5_max_diff} to extract the corresponding  number of $N_{\rm chb}$ terms.  With our parameters we find an acceptable tolerance at about $N_{\rm chb} \simeq 4 \Delta H/(\varepsilon_L-\varepsilon_H)$, but this factor could be somewhat different depending on the application and the location of the gap relative to the spectrum minima.
Our results would have immediate implication on cases such as sGW, where traditional Chebyshev filtering is expensive and the simpler alternative, explicit projection by $\hat{\rho} = \sum_{n\le N_{\rm occ}} |\phi_n\rangle \langle \phi_n|$  (see Ref. ~\cite{vlvcek2018swift}), requires for large systems huge memory  for storing all the occupied eigenstates. 

Finally, our work indicates that when there is at least partial knowledge of the system spectrum, a polynomial expansion could be made more efficient than an expansion which is designed to be uniformly convergent as a function of energy.

\subsubsection*{CRediT authorship contribution statement}
\textbf{Minh Nguyen}: Investigation, Writing - Original
Draft, Software, Visualization. 
\textbf{Daniel Neuhauser}: Conceptualization, Methodology, Supervision, Writing - Review \& Editing, Funding acquisition

\section*{Declaration of Competing Interest}
The authors declare that they have no known competing financial interests or personal relationships financial interests or personal relationships
that could have appeared to influence the work reported in this paper.

\section*{Data Availability Statement}
The data that support the findings of this study are available from the
corresponding author upon reasonable request.

\section*{Acknowledgements}
We are grateful to Nadine Bradbury for useful discussions.  This paper was supported by the Center for Computational Study of Excited State Phenomena in Energy Materials (C2SEPEM), which is funded by the U.S. Department of Energy, Office of Science, Basic Energy Sciences, Materials Sciences and Engineering Division via Contract No. DE-AC02- 05CH11231, as part of the Computational Materials Sciences Program.
Computational resources were supplied through the XSEDE allocation TG-CHE170058, and the NERSC project m4022.

\bibliographystyle{elsarticle-num-names} 
\bibliography{bib}

\end{document}